\documentclass[aip,
amsmath,amssymb,
reprint,%
]{revtex4-1}

\usepackage{graphicx}
\usepackage{dcolumn}
\usepackage{bm}

\usepackage[utf8]{inputenc}
\usepackage[T1]{fontenc}
\usepackage{mathptmx}
\usepackage{etoolbox}
\usepackage{caption}
\usepackage{subcaption}
\usepackage[usenames, dvipsnames]{xcolor} 
\usepackage{soul}
\usepackage{upgreek}
\usepackage{hyperref}
\usepackage{booktabs}
\usepackage{tabularx}
\makeatother
\begin{document}

\preprint{AIP/123-QED}

\title[]{Multifrequency-resolved Hanbury Brown--Twiss Effect}

\author{Joseph Ferrantini}
\altaffiliation[]{These authors contributed equally to this work.}
\affiliation{ 
Brookhaven National Laboratory,
Upton NY 11973, USA 
}%

\author{Jesse Crawford}
\altaffiliation[]{These authors contributed equally to this work.}
\affiliation{ 
Brookhaven National Laboratory,
Upton NY 11973, USA 
}%

\author{Sergei Kulkov}
\altaffiliation[]{These authors contributed equally to this work.}
\affiliation{
Faculty of Nuclear Sciences and Physical Engineering, Czech Technical University, 115 19 Prague, Czech Republic
}

\author{Jakub Jirsa}%
\affiliation{
Faculty of Nuclear Sciences and Physical Engineering, Czech Technical University, 115 19 Prague, Czech Republic
}
\affiliation{
Faculty of Electrical Engineering, Czech Technical University, 166 27 Prague, Czech Republic
}

\author{Aaron Mueninghoff}%
\affiliation{ 
Stony Brook University, Stony Brook NY 11794, USA 
}%

\author{Lucas Lawrence}%
\affiliation{ 
Brookhaven National Laboratory,
    Upton NY 11973, USA 
}%

\author{Stephen Vintskevich}%
\affiliation{ 
Technology Innovation Institute, Abu Dhabi, United Arab Emirates
}%

\author{Tommaso Milanese}%
\affiliation{ 
École polytechnique fédérale de Lausanne (EPFL), CH-2002 Neuchâtel, Switzerland
}%

\author{Samuel Burri}%
\affiliation{ 
École polytechnique fédérale de Lausanne (EPFL), CH-2002 Neuchâtel, Switzerland
}%
\author{Ermanno Bernasconi}%
\affiliation{ 
École polytechnique fédérale de Lausanne (EPFL), CH-2002 Neuchâtel, Switzerland
}%

\author{Claudio Bruschini}%
\affiliation{ 
École polytechnique fédérale de Lausanne (EPFL), CH-2002 Neuchâtel, Switzerland
}%

\author{Michal Marcisovsky}%
\affiliation{ 
Faculty of Nuclear Sciences and Physical Engineering, Czech Technical University, 115 19 Prague, Czech Republic
}%

\author{Peter Svihra}%
\affiliation{ 
Faculty of Nuclear Sciences and Physical Engineering, Czech Technical University, 115 19 Prague, Czech Republic
}%

\author{Andrei Nomerotski}%
\affiliation{
Faculty of Nuclear Sciences and Physical Engineering, Czech Technical University, 115 19 Prague, Czech Republic
}
\affiliation{
Florida International University, Miami FL 33199, USA
}

\author{Paul Stankus}%
\affiliation{ 
Brookhaven National Laboratory,
    Upton NY 11973, USA 
}%

\author{Edoardo Charbon}%
\affiliation{ 
École polytechnique fédérale de Lausanne (EPFL), CH-2002 Neuchâtel, Switzerland
}%

\author{Raphael A. Abrahao}%
 \email{rakelabra@bnl.gov}
\affiliation{ 
Brookhaven National Laboratory,
    Upton NY 11973, USA 
}%

\date{\today}

\begin{abstract}
\textbf{Abstract:} {The Hanbury Brown-Twiss (HBT) effect holds a pivotal place in intensity interferometry and gave a seminal contribution to the development of quantum optics. To observe such an effect, both good spectral and timing resolutions are necessary. Most often, the HBT effect is observed for a single frequency at a time, due to limitations in dealing with multifrequencies simultaneously, halting and limiting some applications. Here, we report a fast and data-driven spectrometer built with a one-dimensional array of single-photon-sensitive avalanche diodes. We report observing the HBT effect for multifrequencies at the same time. Specifically, we observed the HBT for up to 5 lines of the Ne spectrum, but this can be improved even further. Our work represents a major step to make spectral binning and multifrequencies HBT more widely available. The technology we present can benefit both classical and quantum applications.}
\end{abstract}

\maketitle

\section{Introduction}

The Hanbury Brown-Twiss (HBT) effect, i.e., the bunching of photons coming from a thermal light source, became a pivotal effect for intensity interferometry~\cite{HBT_Sirius1956,HBT_photoncorr1956,brown1974intensitybook,eisenhauer2023advances,Monnier2003,brown1957interferometry,brown1958interferometry1,brown1958interferometry2,brown1958interferometry3}, widely used in astronomy, and played a crucial role in the development of quantum optics~\cite{fano1961,glauber1963quantum_optical_coherence,MFox_QObook,milonni2019introductionQO,loudon2000_QObook}. More recently, the HBT effect contributed to the emerging field of quantum astrometry~\cite{stankus2022two,Crawford2023}. Additionally, the HBT found its applications in high energy physics, specially in nuclear and particle collisions~\cite{baym1998HBT}, has been the basis for a plethora of quantum physics experiments~\cite{bromberg2010HBT,magana2016HBT_twistedlight,henny1999fermionicHBT,schellekens2005HBTultracold,oliver1999HBTtype,coloredHBT_2016}, and influenced the development of imaging methods~\cite{PRL2019_optimal_imaging}.

To observe the HBT effect one needs both fine spectral and temporal resolution; poor resolution in either quantity can wash out the visibility of the effect. Here, we report observing the HBT effect for 5 distinct frequencies at the same time, i.e., light comprised of a combination of 5 different frequencies impinging on half of a single-photon-sensitive sensor array in parallel.  One expects to observe the HBT effect, namely an enhancement in the rate of coincident pairs, for photons of the same frequency. Our work is an example of what we call a spectral binning technique, in which different frequencies can coexist in our setup at the same time and each frequency can be measured independently~\cite{jirsa2023fast}; in practice, it allows one to run multiple experiments in parallel. This was achieved by employing our fast and data-driven single-photon sensitive spectrometer based on the LinoSPAD2 detector~\cite{jirsa2023fast,milanese2023linospad2,bruschini2023}, providing simultaneous data collection at multiple wavelengths with good signal-to-noise ratios.

The technology reported here was developed to be used in quantum astrometry~\cite{stankus2022two,Crawford2023,jirsa2023fast}, but it has broad applications in both classical and quantum optics. Some of these include approaches to quantum-enhanced telescopes~\cite{gottesman2012longer,PRL2023_Oregon,Khabiboulline_paper1_PRA,Khabiboulline_paper2_PRL,Marchese&Kok_PRL2023,Czupryniak&Kwiat_PRA2023} and to intensity interferometry~\cite{walter2023resolving,guerin2017temporal,guerin2018spatial,rivet2020intensity,abeysekara2020demonstration,de2022combined,karl2024photon,PRA_Pearce2015,NJP_bojer2022,MNPAS_karl2022comparing}, fluorescence imaging~\cite{lichtman2005fluorescence}, remote sensing~\cite{aasen2018quantitative}, quantum communications~\cite{ciurana2014quantum}, and frequency-bin quantum information~\cite{Lu2023_review}. It can benefit any field requiring high spectral and temporal resolutions simultaneously.

\section{Experimental Setups}

\subsection{LinoSPAD2 detector}

The LinoSPAD2 detector consists of a daughterboard with a sensor of a linear array of 512 Single-Photon Avalanche Diodes (SPAD) and two Field Programmable Gate Array (FPGA) motherboards that read out half of the sensor each. With a pitch of 26.2 $\upmu$m, the whole sensor is approximately 13 mm long. With a fill factor of 57.7\% for a device with microlenses, the photon detection efficiency (PDE) covers essentially the whole visible spectrum and peaks at $\approx30\%$ for approximately 520 nm~\cite{bruschini2023challenges, milanese2023linospad2}. For the Ne spectral lines used for HBT measurements in this work, the PDE goes from 19\% for the Ne 633.4 nm line to 13\% for the 703.2 nm one. A median dark count rate of 100 counts/second (cps) per pixel at room temperature and a bias voltage of 4 V makes it possible to operate the detector in ambient conditions~\cite{bruschini2023}. Cross-talk was measured at $\approx0.2\%$ for the immediate neighbors, falling to $\approx0.01\%$ for further neighbors. The average timing resolution of the detector for a single-photon detection is 40 ps r.m.s.~\cite{jirsa2023fast}. Only one half of the sensor was utilized in this work as 256 pixels were sufficient to fit multiple Ne spectral lines, and operating a single half of the sensor simplifies both readout and analysis. This also removes any need for synchronization between the two halves. Further information on the LinoSPAD2 sensor can be found in Refs.~\onlinecite{jirsa2023fast,milanese2023linospad2}.

\subsection{Single-line setup}

\begin{figure}
  \includegraphics[width = 1.0\linewidth]{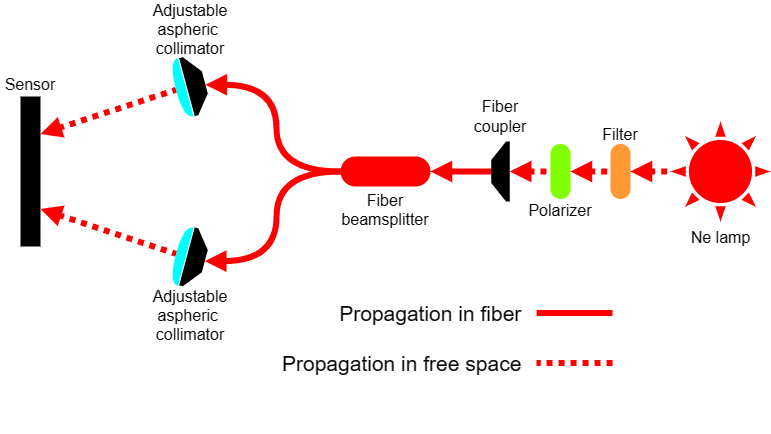}
    \caption{Diagram of the single-line setup with LinoSPAD2. Polarized and filtered Ne light is fiber-coupled to a 1-to-2 50:50 single-mode fiber beamsplitter. The filter is used to select a single line from the Ne spectrum. Each arm of the beamsplitter is connected to an adjustable aspheric collimator, which focuses light onto the LinoSPAD2 sensor.}
    \label{fig:2_fiber_setup_diagram}
 \end{figure}

We begin with the simplest case of observing an HBT measurement for a single atomic line. The experimental setup is conceptually depicted in Fig. \ref{fig:2_fiber_setup_diagram}. As a thermal source of light,  a Ne calibration lamp was used (Newport, model 6032), which was operated at 10 mA AC.  Light from the Ne lamp passes through a spectral filter with the central wavelength of $700$ nm and FWHM of 10 nm, covering the 703.2 nm Ne spectral line with a transparency of 98.7\%. Following, light passes through a linear polarizer, which reduces the detected rate in half, but purifies the polarization state of the light, such that we end up with a twofold increase in the HBT contrast. After that, the light is coupled to a single-mode optical fiber. This fiber is connected to a 1-to-2 50:50 single-mode fiber beamsplitter (Thorlabs TW670R5F1). The outputs of the beamsplitter were collimated using adjustable aspheric collimators, resulting in two illuminated dots, each covering 3 to 5 pixels of the sensor. The results of this configuration are discussed in \ref{Single_Ne_line_HBT}.

\subsection{Dual spectrometer setup}

\begin{figure}
     \includegraphics[width = 1.0 \linewidth]{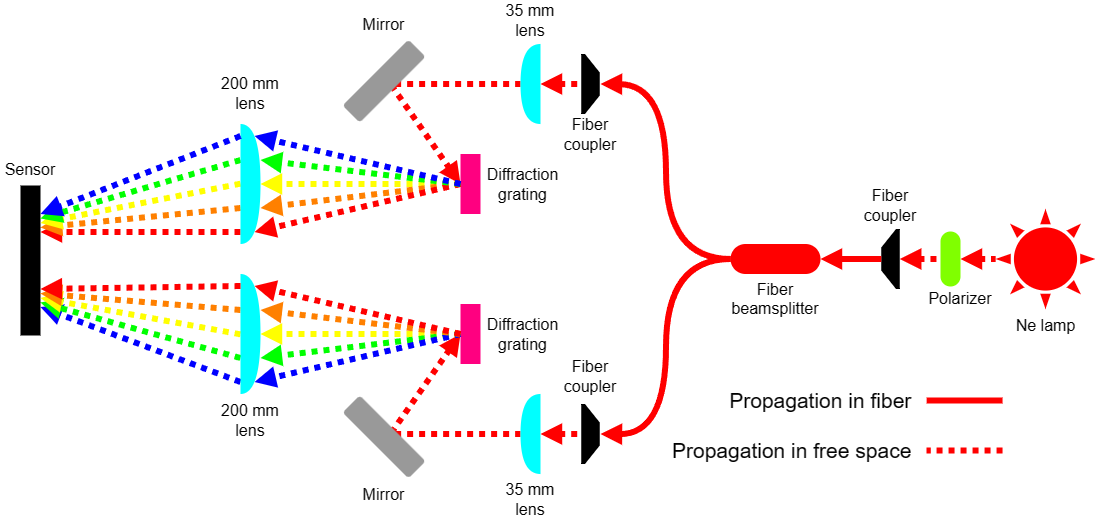}
     \caption{Diagram of the dual spectrometer setup with LinoSPAD2. Light from a Ne lamp is polarized, coupled to a 1-to-2 50:50 single-mode fiber beamsplitter, and directed into the two arms of the spectrometer. Once back to free space propagation, the light is collimated, reflected by a mirror onto a reflective diffraction grating, and finally focused onto the LinoSPAD2 sensor via 200 mm focal length lenses. The top and bottom parts of the spectrometer were designed as mirror-like images of each other.}
     \label{fig:schematics}
 \end{figure}

Following the successful results obtained in the single-line setup, a spectrometer setup was assembled.  Here a Ne calibration lamp was again chosen as the source of light due to the abundance of spectral lines in the region of 500--700 nm, where the PDE of the LinoSPAD2 is the highest. The LinoSPAD2 dual spectrometer layout is shown in Fig. \ref{fig:schematics}. As before a polarizer in front of the lamp was again used to achieve a twofold improvement in the contrast of HBT peaks. 

As in the single-line setup, the Ne light is coupled to a single-mode optical fiber, which is connected to the 1-to-2 50:50 fiber beamsplitter. After that, the light enters the free-space portion of the setup. A 35 mm focal length lens in front of each fiber beamsplitter output collimates the light, resulting in a wider beam diameter of $\approx 4 $ mm. These collimated beams are each reflected by a silver-coated mirror onto a 1-inch square ruled reflective diffraction grating with 1200 grooves/mm and 750 nm blaze wavelength. The diffracted light within each arm of the dual spectrometer is finally focused through 200 mm focal length lenses onto the sensor pixels. The wider beam allows a finer focus on the sensor, thus leading to a better spectral resolution. The measured spectral scale is $\approx$ 0.1 nm/pixel within the studied range. A detailed discussion on measured spectral resolutions can be found at Jirsa \textit{et al.}~\cite{jirsa2023fast}.

Both arms of the dual spectrometer are mirror-like images of each other so that light from each beamsplitter output passes through the same kind of components. Each photon detection is timestamped, enabling offline comparison and analysis of the HBT effects between all combinations of the spectral lines. More details of the spectrometer use of the LinoSPAD2 sensor can be found in Jirsa \textit{et al.}~\cite{jirsa2023fast}. The results of using this configuration are discussed in \ref{resultsHBT_2lines} and \ref{resultsHBT_5lines}. Additionally, when analyzing the data of both the single-line setup and the dual spectrometer setup, we used the Python scientific computing library SciPy~\cite{python_scipy}.

\section{Results and discussion}

\subsection{HBT measurements with a single spectral line}
\label{Single_Ne_line_HBT}

\begin{figure}
     \includegraphics[width = 1.0\linewidth]{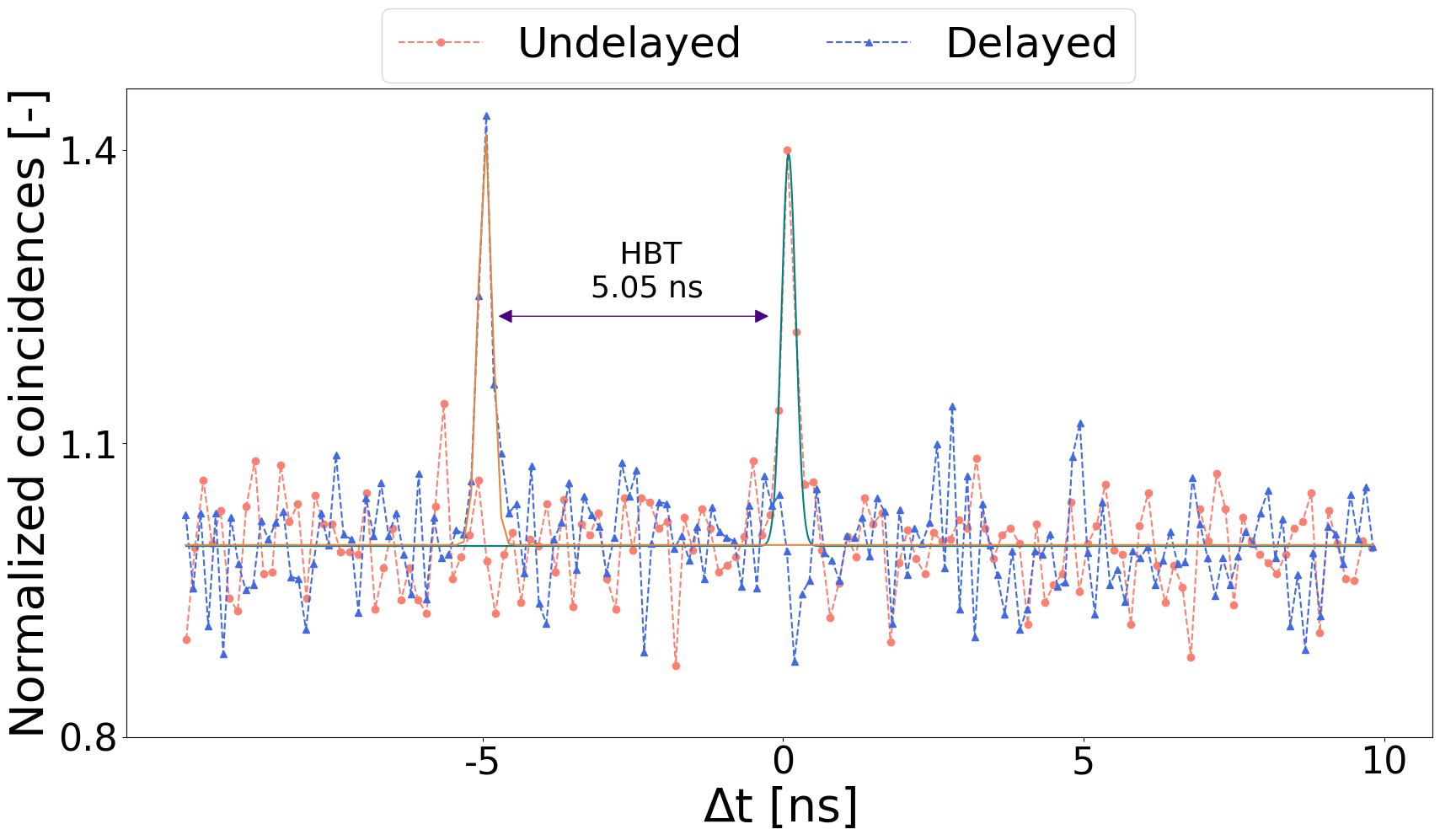}
     \caption{HBT effect with the LinoSPAD2 detector in a single-line setup with a filtered and polarized Ne lamp as a light source. The plot shows photon coincidence counts normalized to the median versus the timestamp difference for the 703.2 nm Ne line for the delayed and undelayed cases. The resulting shift confirms the HBT effect. The average standard deviation is $0.12\pm0.02$ ns, and the average contrast is $41.4\pm6.4 \%$.}
     \label{fig:SK_two_fiber_pix3_45_703nm}
 \end{figure}

To test the capabilities of the LinoSPAD2 to measure the HBT effect, we first used the single-line setup for the 703.2~nm line from the Ne spectrum. Outputs of the two collimators were positioned at pixels 3 and 45, partially to minimize the effect of cross-talk between pixels of interest, and data were recorded for $\approx4$ minutes. As each collimated beam covered approximately 5 pixels each, only the pixels with the highest number of photons detected were used for the HBT analysis. Recorded photon timestamps between the two pixels were compared to calculate the time differences $\Delta t$ between timestamp pairings. The HBT effect predicts an increase in photon coincidences for $\Delta t=0$, which appears as a peak called the HBT peak, the height of which is determined by source coherence time (i.e., spectral width). The HBT effect can be smeared out by the detector time response, which in turn creates a demand for fast detectors. The measured HBT peak is shown in the top part of Fig.~\ref{fig:SK_two_fiber_pix3_45_703nm}. The enhanced peak of coincidence counts --- the HBT effect itself --- is clear and visually well above the noise level.

To further confirm that the enhancement in photon coincidences is indeed due to the HBT effect and not some undesirable effect, i.e., cross-talk between pixels, an additional 1~m single-mode optical fiber was introduced into one of the output arms of the beamsplitter. Considering that the fiber is approximately 1 m long, this should result in a $\approx 5$ ns shift of the HBT peak, and as the cross-talk effect is purely inherent to the detector sensor, it remains unaffected by any changes in the photons paths before they reach the sensor. The HBT peak was indeed confirmed, which can be seen in Fig.~\ref{fig:SK_two_fiber_pix3_45_703nm}. The resulting temporal shift of the peak position is affected by the fiber length and by the non-ideal calibration of the detector. Nevertheless, the experimental evidence is conclusive. The average measured contrast of the HBT peak is $41.4\pm6.4 \%$.

\subsection{HBT measurements with two spectral lines}
\label{resultsHBT_2lines}

\begin{figure}
     \includegraphics[width = 1\linewidth]{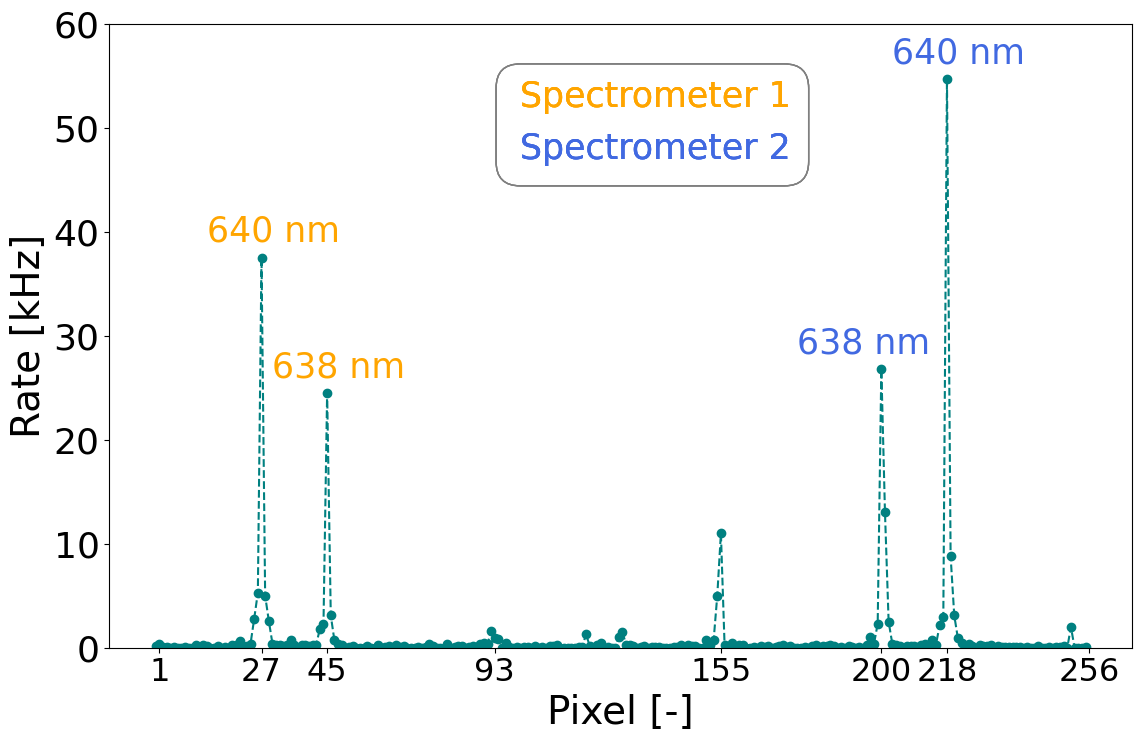}
     \caption{Two copies of part of the Ne spectrum with lines 640.2 nm and 638.3 nm visible. Since the dual spectrometers are mirror-like designs of each other, the wavelength will increase for one arm and decrease for the other with the increasing pixel number. The peak between the two pairs seen at pixel 155 is most probably the 633.4 nm Ne line --- the same line is seen at pixel 93.}
     \label{fig:spectra}
 \end{figure}

\begin{figure}[]
     \includegraphics[width = 1\linewidth]{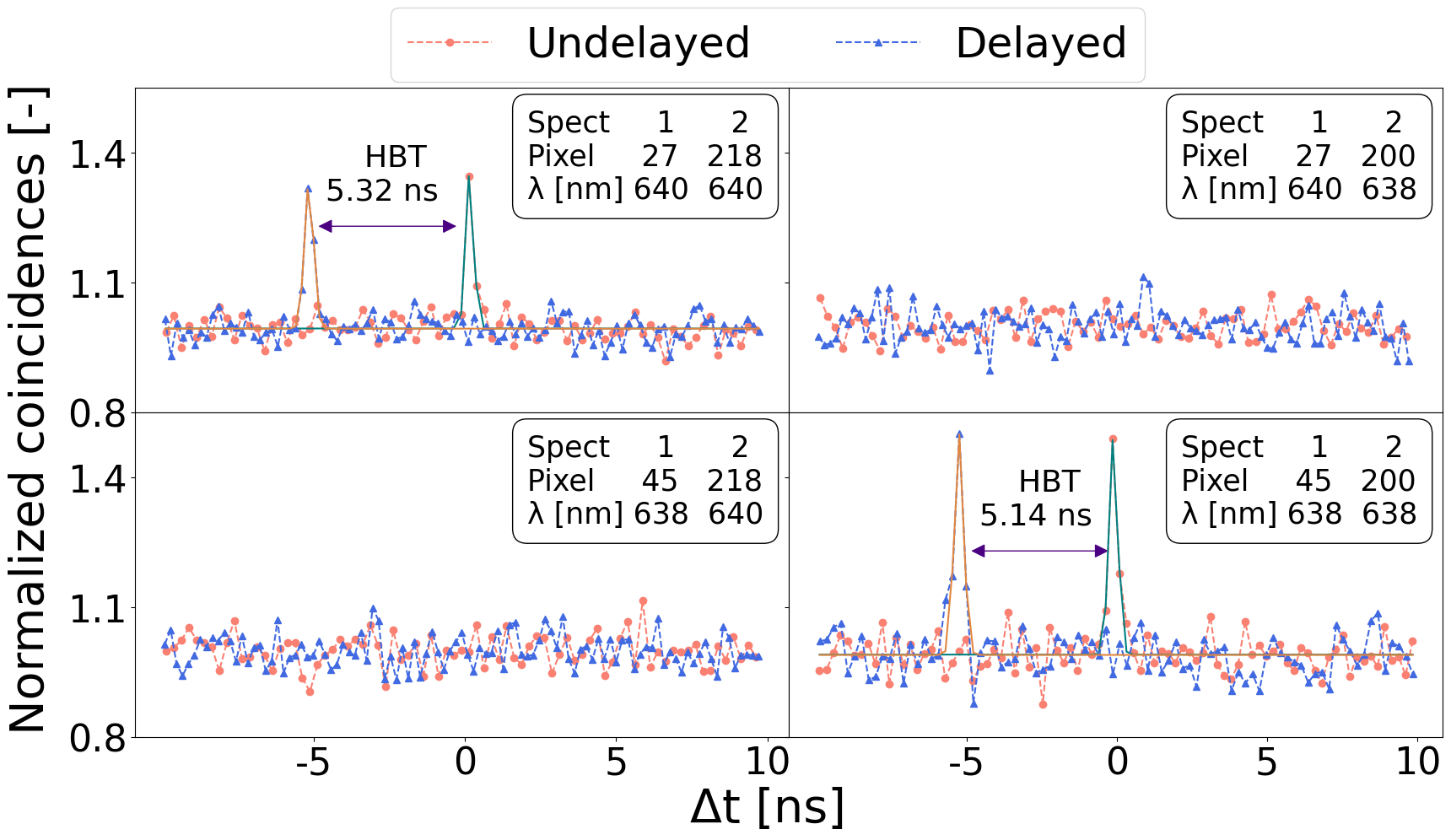}
    \caption{Measurement of HBT visibility for 2 spectral lines. Histograms of time differences between timestamps for different pairings of spectral lines. Both the undelayed and delayed data sets are shown. The bin size is $\approx200$ ps. The top left corner shows the HBT peak for the 640.2~nm line, while the bottom right corner shows the HBT peak for the 638.3~nm line. The average shift due to the added 1 m-long fiber is $5.18\pm 0.03$ ns. The inserts list each line, its position on the sensor, and the corresponding spectrometer arm. The HBT peak contrasts are reported in Table \ref{tab:2x2_fit_params}.}
\label{fig:hbt_peak_comparisons}
\end{figure}

With the HBT effect confirmed for a single pair of lines, we moved to the dual spectrometer setup and tested two pairs of two different Ne spectral lines, namely 638.3 nm and 640.2 nm. Figure \ref{fig:spectra} shows the parts of Ne spectra as recorded by the LinoSPAD2, specifically the 638.3 nm and 640.2 nm spectral lines. Using the two lines seen at pixels 27 and 45, the spectral resolution can be estimated at $\approx 0.1$ nm/pixel. The total data acquisition time for this data set is $\approx30$ min. The different rates observed for different wavelengths are due to the difference in intensities of those Ne lines and for the lines of the same wavelength due to the unequal split of the beamsplitter at the measured wavelengths. All four combinations of 638.3 nm and 640.2 nm lines were analyzed, where HBT is expected only for the pairs of the same wavelength. This is indeed observed,  see Fig. \ref{fig:hbt_peak_comparisons}. Additionally, to verify the effect, a single-mode optical fiber 1 m long was inserted between one output of the beamsplitter and the fiber-to-free-space coupler --- similar to how it was done for the single-line setup. The shift in the HBT peak position was indeed observed with an average of $5.18\pm 0.03$ ns. The fall in the contrast measured between the undelayed and delayed data sets could be due to the decrease in intensity measured for the delayed set, as we hypothesize that the additional 1 m long fiber could have a dusted core, thus leading to a signal loss.

HBT peaks were fitted using the Gaussian function to get the HBT peak position and contrast. The other fitting parameters for all 4 HBT peaks seen in Fig. \ref{fig:hbt_peak_comparisons} are reported in \mbox{Table \ref{tab:2x2_fit_params}}.

\begin{table}[]
	\centering
	\begin{tabular}{c|c|c|c}
		\toprule
		Pixel pair & Wavelength [nm] & Peak position [ns] & Contrast [\%] \\
		\midrule
        27, 218 & 640.2 & \phantom{0} $0.16 \pm 0.02$ & $ 36.9 \pm 9.6$ \\
        45, 200 & 638.3 & $-0.11 \pm 0.02$ & $ 50.9 \pm 8.0$ \\
		\midrule
        27, 218 & 640.2 & $-5.16 \pm 0.01$ & $ 34.0 \pm 8.4$ \\
        45, 200 & 638.3 & $-5.25 \pm 0.02$ & $ 51.6 \pm 9.0$ \\
		\bottomrule
	\end{tabular}
	\caption{Fitting parameters for the HBT with 2 spectral lines. The top two rows are for the undelayed dataset, and the bottom two are for the delayed one. The average standard deviation is $0.15\pm0.02$ ns.}
	\label{tab:2x2_fit_params}
\end{table}

\subsection{HBT measurements with five spectral lines}
\label{resultsHBT_5lines}

Next, we moved to the situation where we observed HBT for five pairs of Ne lines. In this case, we are clearly in the regime of multifrequency-resolved HBT.

\begin{figure}
    \includegraphics[width=1\linewidth]{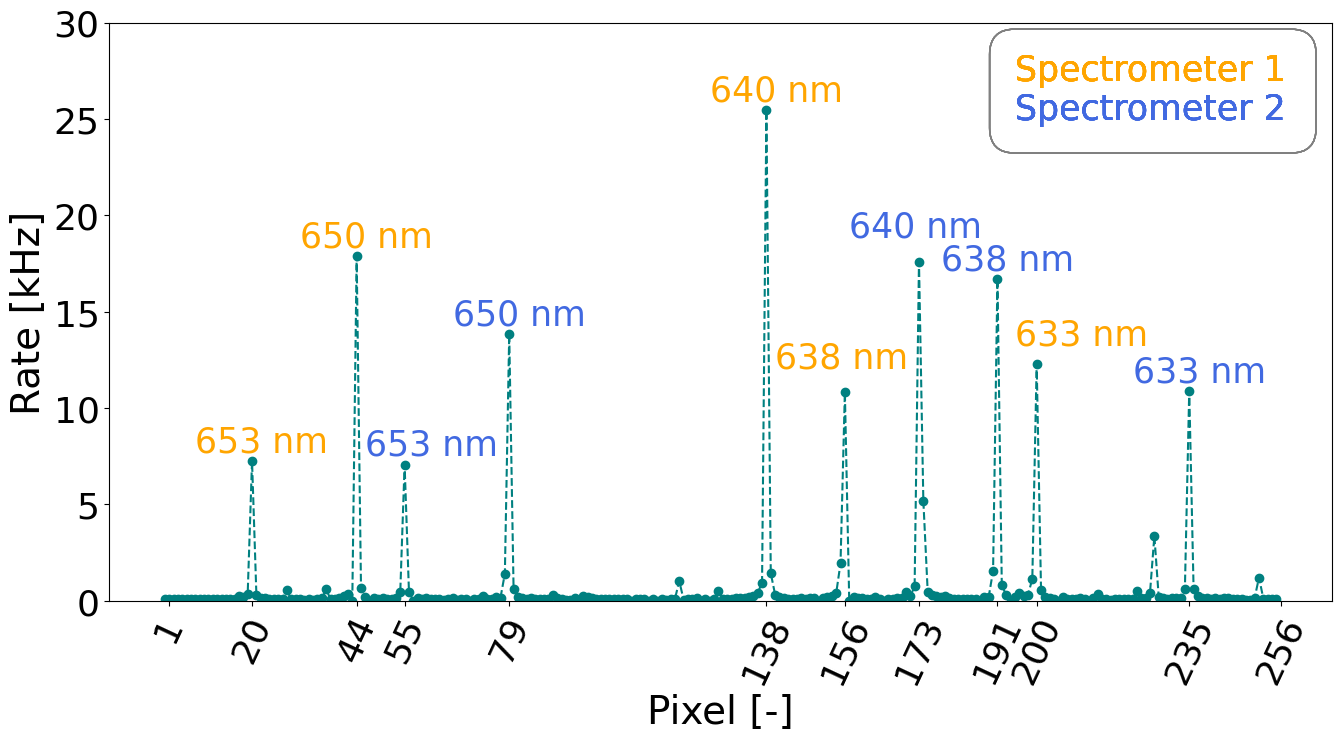}
    \caption{Two copies of a five-line portion of the Ne spectrum, side by side, as detected by the LinoSPAD2 from the dual spectrometer setup. In the new vertical orientation of the sensor, the wavelengths for both spectrometer arms will decrease as the pixel number increases.}
    \label{fig:5_line_senpop1}
\end{figure}

The biggest limitation of the linear sensor of LinoSPAD2 is the difficulty of alignment as the sensor is not only single-pixel wide but also relatively narrow given the pixel size of $26.2 \ \upmu\mathrm{m}$. Because of that, even with a spectral scale of 0.1 nm/pixel, it was possible to focus only two Ne spectral lines from each arm of the dual spectrometer. By reorienting the LinoSPAD2 sensor from a horizontal to a vertical orientation with respect to the ground and adjusting the positions and orientations of mirrors and gratings accordingly, alignment was successfully facilitated. As a result, it was possible to focus 10 Ne lines, 5 from each arm of the dual spectrometer. The resulting spectra are presented in Fig. \ref{fig:5_line_senpop1}. Similar to the two-line spectrometer setup, the difference in the photon rates for the lines of the same wavelength can be accounted for by the non-equal split of the 1-to-2 beamsplitter. Apart from the orientation of the components and the sensor, the setup used here is the same as was used for the measurements with two pairs of Ne spectral lines. Therefore, the spectral scale is still 0.1 nm/pixel.

\begin{figure*}
    \includegraphics[width=1\linewidth]{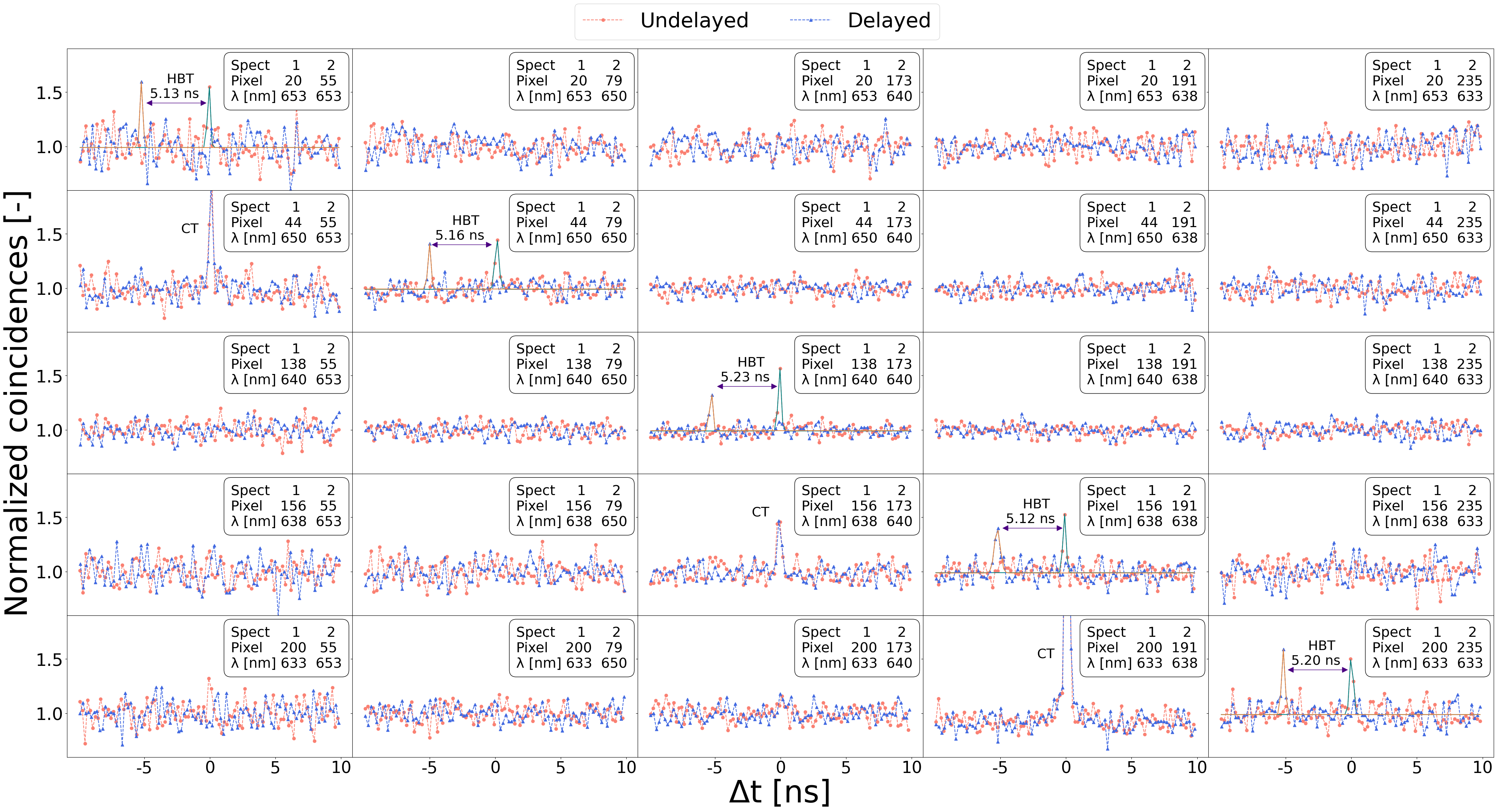}
    \caption{HBT effect for 5 spectral lines. Histograms of time differences between all combinations of 5 pairs of different Ne spectral lines. Histograms along the diagonal, starting from the top left, show both shifted and unshifted HBT peaks. The shifted peaks were all delayed by $5.17\pm 0.03$ ns on average due to inserting a 1 m fiber into one arm of the spectrometer. Additionally, cross-talk peaks can be also seen for three different pairs of not-too-distant pixels. The fact that the addition of a 1~m fiber to one of the spectrometers shifts the peaks seen in the pairs of the same wavelength and not for the cross-talk peaks further confirms the HTB effect. The inserts list each line, its position on the sensor, and the corresponding spectrometer arm. The HBT contrasts are reported in Table~\ref{tab:5x5_fit_params}.}
    \label{fig:5x5_delta_t}
\end{figure*}

Histograms of timestamp differences for all combinations of pairs of lines can be seen in Fig. \ref{fig:5x5_delta_t}. The HBT peaks are seen only for the pairs of lines of the same wavelength and are again verified via the addition of a 1 m long optical fiber. Additionally, three cross-talk peaks can be seen. These cross-talk peaks provide further proof of the HBT effect as they do not shift after adding the 1 m long fiber to one of the spectrometer arms. The difference in shifts of the HBT peaks between the different pair combinations can be accounted for by the nonideal offset calibration, so that the undelayed HBT peaks do not always appear at $\Delta t = 0$, and that offset is different for different pairs of pixels. The average shift is $5.17\pm 0.03$ ns. All HBT peaks were fitted using the Gaussian function, and the average standard deviation is $0.13\pm0.02$ ns, while the other fit parameters can be seen in Tab. \ref{tab:5x5_fit_params}. 

In principle, the uncertainties could be reduced by taking longer datasets. Most importantly, the evidence for the observation of the multifrequency-resolved HBT effect is undeniable, and in most cases, the HBT contrast is good enough to enable further experiments.

\begin{table}[]
	\centering
	\begin{tabular}{c|c|c|c}
	\toprule
	Pixel pair & Wavelength [nm] & Peak position [ns] & Contrast [\%] \\
	\midrule
	44, 79 & 650.7 & $ -0.08 \pm 0.08$ & $ 62.8 \pm 51.7$ \\
	20, 55 & 653.3 & \phantom{0} $ 0.13 \pm 0.03$ & $ 46.7 \pm 15.8$ \\
	138, 173 & 640.2 & $ -0.07 \pm 0.02$ & $ 59.9 \pm 10.2$ \\
	156, 191 & 638.3 & $ -0.11 \pm 0.02$ & $ 54.7 \pm 12.9$ \\
	200, 235 & 633.4 & \phantom{0} $ 0.01 \pm 0.03$ & $ 56.6 \pm 20.5$ \\
	\midrule
	44, 79 & 650.7 & $ -5.21 \pm 0.07$ & $ 61.5 \pm 24.6$ \\
	20, 55 & 653.3 & $ -5.02 \pm 0.05$ & $ 42.9 \pm 18.7$ \\
	138, 173 & 640.2 & $ -5.30 \pm 0.04$ & $ 40.4 \pm 45.4$ \\
	156, 191 & 638.3 & $ -5.23 \pm 0.05$ & $ 40.8 \pm 17.6$ \\
	200, 235 & 633.4 & $ -5.18 \pm 0.03$ & $ 60.3 \pm 13.7$ \\
	\bottomrule
	\end{tabular}
	\caption{Fitting parameters for the HBT with 5 spectral lines. The top five rows are for the undelayed dataset, and the bottom five are for the delayed one. The average standard deviation is $0.14\pm0.03$ ns.}
	\label{tab:5x5_fit_params}
\end{table}

\section{Conclusion}

The Hanbury Brown-Twiss effect was not only important for its contribution to understanding the nature of light and quantum optics, but also found a vast set of applications, especially in intensity interferometry, astronomy, and astrophysics. 

In the present work, we showed how we achieved HBT with multiple frequencies at the same time. We first demonstrated the HBT effect using a single line of Ne spectrum, then 2 lines, and lastly with 5 lines. We observed and confirmed the HBT effect for photons of the same wavelength, and no HBT effect for different wavelengths, as theory predicts.

We used a spectrometer setup that we built using a one-dimensional array of 256 pixels of the LinoSPAD2 sensor, with single-photon sensitivity in each pixel. We could extend the work beyond 5 lines, there is no fundamental limitation preventing that. However, to achieve that, a larger sensor array would be necessary. Moreover, a detector with similar timing capabilities but with a 2D sensor matrix would facilitate the optical alignment. This would allow for the use of echelle gratings, potentially improving spectral resolution. 

Another experiment~\cite{coloredHBT_2016} reports observing HBT for various frequencies, where a polariton condensate that radiates as a spectrally broad light source was utilized. Additionally, Refs.~\onlinecite{Grabarnik2008,Jennewein_spectr2014SR,Jennewein_spectr2014JAP, Lubin2021,Farella_AIP2024,morimoto2020megapixel} report on other 1D and 2D spectrometers.

In our experiment, the entire setup, from the light source to the detector, operates at room temperature, which simplifies the use of our technology, as well as mitigates the costs of scaling. The ease of use and the high temporal and spectral resolution of the LinoSPAD2 dual spectrometer setup we developed can be highly beneficial for classical and quantum applications.

Our aim is to apply this technology for quantum astrometry~\cite{Crawford2023}. Other approaches of quantum-enhanced telescopes~\cite{gottesman2012longer,PRL2023_Oregon,Khabiboulline_paper1_PRA,Khabiboulline_paper2_PRL,Marchese&Kok_PRL2023,Czupryniak&Kwiat_PRA2023} could also benefit from this technology, as well as classical intensity interferometers. In fact, this technology is applicable to any experiment interested in spectral binning.

\begin{acknowledgments}

\end{acknowledgments}

This work was supported by the U.S. Department of Energy QuantISED award, the Brookhaven National Laboratory LDRD grant 22-22, the Ministry of Education, Youth and Sports of the Czech Republic Grant No. LM2023034, as well as from European Regional Development Fund-Project ``Center of Advanced Applied Science" No. CZ.02.1.01/0.0/0.0/16-019/0000778. This work was also supported by the EPFL internal IMAGING project ``High-speed multimodal super-resolution microscopy with SPAD arrays", the DOE/LLNL project ``The 3DQ Microscope", and the Grant Agency of the Czech Technical University in Prague, grant No. SGS24/063/OHK4/1T/14.
\\

\section*{Data Availability Statement}

The data that support the findings of this study are available from the corresponding author upon reasonable request.

\section*{Author Declaration}
The authors have no conflicts to disclose.

\section*{References}
\bibliography{aipsamp}

\end{document}